\documentclass[prl,twocolumn,showpacs]{revtex4}  
\usepackage{graphicx}  
\usepackage{epstopdf}

\newcommand{\Sinh}{sh}

\begin{document}


\title{Asymptotic Theory of Channeling in the Field of an Atomic Chain and 
an Atomic Plane
\footnote{Soviet Physics Solid State, v27, No.6,  pp.1007-1012  (1985)} }

\author{G.\,V.\, Kovalev \/\thanks}



\affiliation{Moscow, USSR}

\date{Submitted July 23, 1984; resubmitted November 19, 1984}

\begin{abstract}
A rigorous theory of diffraction scattering from extended objects is proposed. The present theory is based on a 
multiple asymptotic expansion of an integral equation for the exact wave function in terms of the large 
parameters of the problem, which are the range of the potential and the momentum components of the 
incident particle. For small angles of incidence the density of positively charged particles on the axis of a 
chain is always lower than unit y and the density of negatively charged particles has a maximum for certain 
strength of the potential which can exceed considerably unity. The conditions of validity of the proposed 
approach are obtained. 
\end{abstract}

\maketitle


\section{1. Introduction}
\label{sec:introduction}

The present theory of channeling is based on the assumption that the potential of a set of atoms forming a 
crystal can be replaced by the potential of a system of 
continuous atomic chains or planes. This approach, first 
proposed by Lindhard [1], can be regarded as a model approach. Its quantum-mechanical version [2 ,3] predicts a number of effects associated with channeling. In particular, radiation which accompanies channeling has received considerable attention [4 - 6].  Although the model approach is widely used, it is not quite satisfactory since, strictly speaking, it is not an approximate method and, therefore, it cannot describe correctly finer effects which require extension beyond the average potential such as 
discreteness of the potential, boundary effects, interstitial atoms, vacancies, etc. The average potential in 
the aforementioned model approach is introduced artificially (as the zeroth Fourier component of the real potential) and longitudinal scattering disappears in this approach. It follows that the problem of scattering from a 
realistic potential bounded in space cannot be treated by this method.

It is our aim to study elastic scattering of fast particles from the potentials of an atomic chain and an 
atomic plane within the general scattering theory. It will 
be shown quite rigorously that the average potential of 
an atomic chain introduced by Lindhard [1] is obtained from 
double asymptotic expansion of the Lippmann-Schwinger 
equation in terms of the length of the atomic chain $L_z \rightarrow \infty$ 
 and in terms of the longitudinal momentum component 
of the incident particle $p_z  \rightarrow \infty$ . It is then possible to formulate a general method for calculating corrections to the zeroth approximation corresponding to the average 
potential since such corrections are simply the next 
higher-order terms in the asymptotic expansion in terms 
of large parameters in the problem.  In the case of an atomic plane, it is necessary to carry out an asymptotic 
expansion of higher multiplicity. The zeroth-order wave 
function is expressed in the present approach in terms of 
a transverse integral equation with a kernel depending on 
the longitudinal length of an atomic chain or an atomic 
plane. An analysis of this equation indicates that the density of the wave function for negatively charged particles 
depends logarithmically on the length of the atomic chain 
near the maximum and can be much greater than unity. 
For real single crystals, this effect is known as flux peaking effect[7].

\section{2. SCATTERING FROM AN EXTENDED ATOMIC CHAIN }
\label{sec:scattering}

We shall consider scattering of fast charged particles 
from the potential of an atomic chain containing $N_z$ atoms 
separated by the lattice period $c$. To ensure that the opposite ends of the chain tend uniformly to $+\infty$  and  $-\infty$ in the limit $N_z \rightarrow \infty$, we shall choose the origin at $z = N_z c/2$. 
The Fourier transform of the potential then remains the same for even and odd $N_z$, and we shall quote the potential  for an odd number of atoms 

\begin{eqnarray}
U({\bf r})=\sum_{j_z=-\frac{N_z-1}{2}}^{\frac{N_z-1}{2}}  U_{at} ({\bf r}+{\bf n_z}  c j_z ), \label{r1a}
 \\
U({\bf k})= U_{at} ({\bf k})  \frac{\sin(N_z k_z \frac{c}{2})}{\sin(k_z \frac{c}{2})}.
\label{r1b}
\end{eqnarray}
Without loss of generality, we shall consider the scattering 
of nonrelativistic particles whose wave function satisfies 
the Lippmann -Schwinger equation ($\hbar = 1$) 
\begin{eqnarray}
 \Psi({\bf r})=e^{i {\bf p r}}+ \int d {\bf r'} \frac{e^{i p |{\bf r-r'}|}}{-4 \pi |{\bf r-r'}|} V({\bf r' }) \Psi({\bf r'}),
\label{r2}
\end{eqnarray}
where  $V({\bf r })= 2 M U({\bf r})$, and $M$ is the particle mass. It is 
well known that the solution of the integral equation (3) 
can be written formally as an infinite Born series 
\begin{eqnarray}
 \Psi({\bf r})=\sum_{n=0}^{\infty} \Psi^{(n)}({\bf r}) 
\label{r3}
\end{eqnarray}
with terms $ \Psi^{(0)}=e^{i {\bf p r}}$, 
\begin{eqnarray}
  \Psi^{(n)}({\bf r}) =e^{i {\bf p r}} \int d {\bf R_1...R_n} \frac{e^{i pR_1- i{\bf p R_1}|}}{-4 \pi R_1}\\ \nonumber
	\times V({\bf r -R_1 }) ...  \frac{e^{i pR_1- i{\bf p R_n}|}}{-4 \pi R_n} V({\bf r -R_1- ...-R_n }) .
\label{r5}
\end{eqnarray}
An eikonal expression for the wave function is obtained 
from the expansion of $ \Psi^{(n)}({\bf r})$ in powers of $p_z \rightarrow \infty$ followed 
by summation of the first terms of this expansion [8]. An 
additional expansion in powers of $L_z \rightarrow \infty$  was considered 
in [9] within the eikonal approximation, i.e.,  a double 
asymptotic expansion has been used (first expansion in 
powers of $p_z \rightarrow \infty$  and then in powers of $L_z \rightarrow \infty$ ). However, the eikonal approximation is not uniform with respect to the length of the potential and holds for $L_z << p_z R^2$. 
It follows that an expansion where the limit $L_z \rightarrow \infty$ is 
taken first and then the limit $p_z \rightarrow \infty$ should yield a different result. In fact, the pole of the Green function (in the momentum space shifted by an amount equal to the particle momentum p) 
\begin{eqnarray}
\frac{1}{k^2+ 2 {\bf p k}- i \delta }, \nonumber
\end{eqnarray}
i.e., the point $k_z =- p_z +  \sqrt{p^2_x - k_{\perp} - 2p^2_x   k_{\perp} - i \delta}$ merges 
in the limit $p_z \rightarrow \infty$  with a removable singularity of the component $k_z = 0$ of the Fourier potential of the chain defined by Eq. (2) and the resulting nonuniform behavior in 
the asymptotic expansion of the integrals involved is due 
to merging of certain critical singularities [10 ,11]. To obtain the wave function in the field of a chain with $L_z \gtrsim p_z R^2$, it is necessary to derive an asymptotic expansion 
of  $\Psi^{(n)}({\bf r})$ which is uniform over the length of the chain $L_z$. 
The parabolic approximation of Leontovich and Fock [12] 
represents such expansion. For $L_z >> p_z R^2$, the parabolic 
approximation can be simplified. The simplification in 
question can be obtained directly from Eq. (5) by expanding with respect to $L_z  \rightarrow \infty$ and then with respect to $p_z  \rightarrow \infty$, and the results are quoted below. Omitting the intermediate steps, we can write the principal term for $\Psi^{(1)}$ in the form 
\begin{eqnarray}
  \Psi^{(1)}({\bf r}) =e^{i {\bf p r}} \int d {\bf R_{1 \perp}} -\frac{i}{8}[H_0^{(1)}(p_{\perp} R_{1 \perp})  \\ \nonumber
+ H_0( \ln \frac{N_z c p_{\perp}}{2 p_z R_{1 \perp}}, p_{\perp} R_{1 \perp})	]  e^{- i{\bf p_{\perp} R_{1 \perp}}}  V_{\perp}({\bf r_{\perp} -R_{1 \perp} },  0) ,
\label{r6}
\end{eqnarray}
Here, $H_0^{(1)} ( \rho )$ is a Hankel function of first kind; $H_0(\beta, \rho)$
is an incomplete Hankel cylindrical function [13]. The quantity $V_{\perp}({\bf r_{\perp} -R_{1 \perp} },  0) $ is the average potential of an atomic chain which is usually introduced to describe the channeling [1].

Since the integrand in Eq.(6) is independent of $r_z$, we can 
introduce quite consistently in Eq. (5) analogous expansions in each of the integrals involved. We obtain 
\begin{eqnarray}
 \Psi^{(n)}({\bf r}) =e^{i {\bf p r}} \int d {\bf R_{1 \perp}}...d {\bf R_{n  \perp}} \tilde{G}_{p_{ \perp}}({\bf  R_{1  \perp}}) e^{- i{\bf p R_{1  \perp}}}   \nonumber  \\ \times V_{\perp} ({\bf r_{\perp} -R_{1 \perp} }, 0) 
	 ... \tilde{G}_{p_{ \perp}}({\bf  R_{n  \perp}}) e^{- i{\bf p R_{n  \perp}}} \nonumber \\ \times V_{\perp} ({\bf r_{\perp} -R_{1 \perp} -... - R_{n \perp} }, 0 ),  
\label{r7}
\end{eqnarray}

\begin{eqnarray}
\tilde{G}_{p_{ \perp}}({\bf  R_{ \perp}})= -\frac{i}{8}[H_0^{(1)}(p_{\perp} R_{\perp})  
+ H_0( \ln \frac{L_z  p_{\perp}}{2 p_z R_{\perp}}, p_{\perp} R_{\perp})	] .
\label{r8}
\end{eqnarray}
Summation of the terms in Eq. (7) yields the wave function in the field of an atomic chain 
\begin{eqnarray}
 \Psi({\bf r})=e^{i  p_{z} r_{z} } \phi_{{\bf p_\perp}} ({\bf r_{\perp} }) ,
\label{r9}
\end{eqnarray}
where $\phi_{{\bf p_\perp}} ({\bf r_{\perp} }) $ satisfies the following two-dimensional 
integral equation: 
\begin{eqnarray}
\phi_{{\bf p_\perp}}=e^{i {\bf p_{\perp} r_{\perp} }}+ \int d {\bf r'_{\perp} } \tilde{G}_{p_{ \perp}}({\bf  r_{ \perp}-r'_{ \perp}}) V_{\perp}({\bf r'_{\perp}},0) \phi_{{\bf p_\perp}} ({\bf r'_{\perp} }). \;\;
\label{r10}
\end{eqnarray}
The solution of the Fredholm Integral equation (10) of 
second kind can be written in the form 
\begin{eqnarray}
 \phi_{{\bf p_\perp}} ({\bf r_{\perp} }) =e^{i {\bf p_{\perp} r_{\perp} }}-\sum_{k=1}^{\infty}\frac{a_k}{1- \lambda_k} \phi^{(k)}_{{\bf p_\perp}} ({\bf r_{\perp} }),
\label{r11}
\end{eqnarray}
where $\lambda_k$ and $\phi^{(k)}$ are the eigenvalues and eigenfunctions 
of the homogeneous Fredholm equation of second kind 
\begin{eqnarray}
 \phi^{(k)}_{{\bf p_\perp}} ({\bf r_{\perp} })=\lambda_k  \int d {\bf r'_{\perp} } \tilde{G}_{p_{ \perp}}({\bf  r_{ \perp}-r'_{ \perp}}) V_{\perp}({\bf r'_{\perp}},0) \phi^{(k)}_{{\bf p_\perp}} ({\bf r'_{\perp} })      \nonumber \\
a_k=\int d {\bf r_{\perp} } e^{i {\bf p_{\perp} r_{\perp} }} V_{\perp}({\bf r_{\perp}},0) \phi^{(k)}_{{\bf p_\perp}} ({\bf r_{\perp}}).  
\label{r12}
\end{eqnarray}
The spectral equation (12) determines the transverse 
eigenfunctions and transverse eigenenergies for channeling in the field of an isolated atomic chain. The eigenfunctions and eigenvalues depend not only on the longitudinal momentum of the particle $p_z$, but also on the longitudinal length of the chain $L_z$.   For an axially symmetric chain, the propagator defined by Eq. (8) can be conveniently written in the form ($\lambda = L_z/4 p_z$) 
\begin{eqnarray}
  -\frac{i}{8}[H_0^{(1)}(p_{\perp} |{\bf r_{\perp}-r'_{\perp}}|)  
+ H_0( \ln \frac{2 \lambda  p_{\perp}}{|{\bf r_{\perp}-r'_{\perp}}|}, \nonumber \\ p_{\perp}|{\bf r_{\perp} - r'_{\perp}}|)	] = \sum_{m=-\infty}^{\infty}e^{i m \varphi}  \tilde{G}^{m}_{p_{ \perp}}( r_{ \perp},r'_{ \perp}), 
\label{r13}
\end{eqnarray}
where $ \varphi$ is the angle between  ${\bf r_{\perp}}$  and ${\bf r'_{\perp}}$,  and 
\begin{eqnarray}
 \tilde{G}^{m}_{p_{ \perp}}( r_{ \perp},r'_{ \perp})=\frac{1}{2 \pi}  \int_{0}^{\infty} k_{\perp} d k_{\perp}  J_{m}( k_{\perp}  r'_{\perp}) J_{m}( k_{\perp}  r_{\perp}) \nonumber \\  \times \frac{1-e^{-i \lambda (k^{2}_{\perp} -p^{2}_{\perp} ) } }{p^{2}_{\perp}-k^{2}_{\perp}+i \delta}.
\label{r14}
\end{eqnarray}
The wave function $ \phi_{{\bf p_\perp}}$ can be now expanded in terms of 
the azimuthal momentum ( $ \varphi_{1}$ is the angle between    ${\bf r_{\perp}}$  and ${\bf p_{\perp}}$) 
\begin{eqnarray}
\phi_{{\bf p_\perp}} ({\bf r_{\perp} }) =\sum^{\infty}_{-\infty} i^m e^{i m \varphi_{1}} \Psi_m( p_{ \perp},r_{ \perp})
\label{r15}
\end{eqnarray}
where $ \Psi_m( p_{ \perp},r_{ \perp})$ satisfies the following integral equation: 
\begin{eqnarray}
\Psi_m( p_{ \perp},r_{ \perp})=J_{m}( p_{\perp}  r_{\perp})  + \nonumber \\ 2 \pi \int^{\infty}_{0} r'_{ \perp} d r'_{ \perp} \tilde{G}^{m}_{p_{ \perp}}( r_{ \perp},r'_{ \perp}) V(r'_{\perp}, 0) \Psi_m( p_{ \perp},r'_{ \perp}).
\label{r16}
\end{eqnarray}
We shall consider two cases corresponding to two limiting values of the parameter $\rho sh(\beta)$ for which the incomplete cylindrical function $H_0( \beta, \rho) $ can be expanded in a series[13].  In the first case when the condition  $\rho *sh(\beta) >>1$ is satisfied and 
\begin{eqnarray}
 H_0( \beta, \rho) = H_0^{(1)}(\rho)+ O(|\rho *  \Sinh(\beta) |^{-1}), \nonumber 
\end{eqnarray}
the kernel defined by Eq. (8) becomes a two-dimensional  Green function $-(i/4)H_0^{(1)} ( p_{ \perp}  r_{ \perp} )$ and Eq. (10) reduces to the usual two-dimensional Lippmann -Schwinger equation independent of $L_z$. In the second case when the inequality  $\rho *sh(\beta) <<1$ holds, we obtain [13]
\begin{eqnarray}
 H_0( \beta, \rho) =\frac{2}{i \pi} \beta+ O(|\rho *  \Sinh(\beta) |). \nonumber 
\end{eqnarray}
This case corresponds to the condition 
\begin{eqnarray}
| \frac{p^2_{\perp} L_z}{4 p_z} - \frac{ |{\bf r_{\perp}-r'_{\perp}}|^2  p_z}{L_z} | <<1. \nonumber 
\end{eqnarray}

\begin{figure}
\centering
		\includegraphics[width=0.45\textwidth]{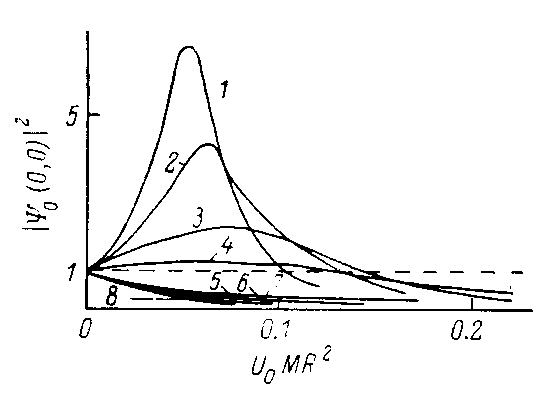}
\caption{Dependencies of the probability density of finding particles on the 
atomic chain axis on the magnitude of the potential $U_0 = 2 Z_1 Z_2 e2/c$ (in 
units of  $[M R^2]^{-1}$ ).  Curves 1 and 8 correspond to scattering of negatively and 
positively charged particles for the following values of the parameter $ \ln(L_z/\gamma p_z R^2)$: 1, 8 - 6.0; 2, 7 - 5.0; 3, 6 - 4.0; 4,5 - 3.0.}
	\label{fig:fig_1}
\end{figure}

i.e., to small angles of incidence $\theta_0 << 2 / \sqrt{ p_z L_z}$ and to 
small distances from the axis of the chain $ |{ \bf r_{ \perp} -r'_{ \perp}}| << \sqrt{L_z / p_z}$. It will be shown that Eq. (10) holds for angles of incidence  $\theta_0  \lesssim 1 / p_z R$  and, therefore, for all points 
within the range of the potential of the chain $|{ \bf r_{ \perp} -r'_{ \perp}}| \lesssim R$
 the condition  $ p_{\perp}|{ \bf r_{ \perp} -r'_{ \perp}}| << 1$ is satisfied and the Hankel function $H_0^{(1)}(p_{\perp} |{ \bf r_{ \perp} -r'_{ \perp}}|)$ can be expanded in a series.  The function $ \tilde{G}_{p_{ \perp}}( { \bf r_{ \perp} })$ then satisfies the following equation: 
\begin{eqnarray}
 \tilde{G}_{p_{ \perp}} \sim \frac{1}{4 \pi} \ln \frac{|{ \bf r_{ \perp} -r'_{ \perp}}|^2  \gamma  p_z}{L_z}- \frac{i}{8} , \;\;   \gamma=1.78... \; 
\label{r17}
\end{eqnarray}
and Eq. (10) can be written in the form 
\begin{eqnarray}
\phi_{{\bf p_\perp}}=e^{i {\bf p_{\perp} r_{\perp} }}+\nonumber \\  \int d {\bf r'_{\perp} } [\frac{1}{4 \pi} \ln \frac{|{ \bf r_{ \perp} -r'_{ \perp}}|^2  \gamma  p_z}{L_z}- \frac{i}{8} ] V_{\perp}({\bf r'_{\perp}},0) \phi_{{\bf p_\perp}} ({\bf r'_{\perp} }).
\label{r18}
\end{eqnarray}
For a potential with azimuthal symmetry, Eqs. (16) and 
(17) yield the foUowing equation for the wave function with 
a momentum $m$:
\begin{eqnarray}
\Psi_m( p_{ \perp},r_{ \perp})=J_{m}( p_{\perp}  r_{\perp})  + \nonumber \\ 2 \pi \int^{\infty}_{0} r'_{ \perp} d r'_{ \perp} \tilde{G}^{'}_{m}( r_{ \perp},r'_{ \perp}) V(r'_{\perp}, 0) \Psi_m( p_{ \perp},r'_{ \perp}).
\label{r19}
\end{eqnarray}

\begin{eqnarray}
  \tilde{G}^{'}_{m}( r_{ \perp},r'_{ \perp}) =\left \{\begin{array}{ll}- \frac{1}{4 \pi m} r^{m}_{\perp <}\; r^{-m}_{\perp >} , &  \; m> 0, \\  
\frac{1}{4 \pi } \ln \frac{r^{2}_{\perp >}   \gamma p_z}{L_z} - \frac{i}{8},&  \;  m=0.
\end{array} \right.
\label{r20}
\end{eqnarray}

\begin{figure}
\centering
		\includegraphics[width=0.45\textwidth]{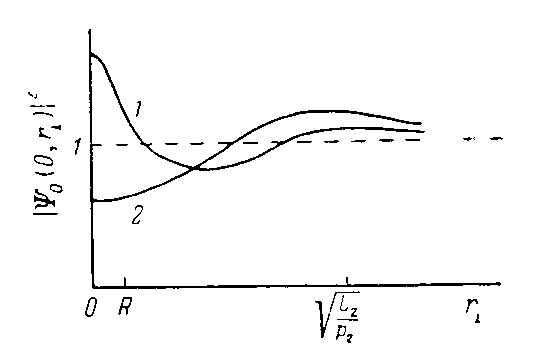}
\caption{Dependencies of the distribution of the particle density on the distance from the axis of an extended chain in the case of the maximum defined by Eq. (23) for $\ln(L_z/\gamma p_z R^2) = 3.2$; 1) negatively charged particles; 2) positively charged particles.}
	\label{fig:fig_2}
\end{figure}

We shall now determine the probability density that 
the partic1e lies on the axis of the chain $r_{\perp} =0$. For $m =0$, Eq. (19) yields 

\begin{eqnarray}
\Psi_0(0,0)=1  + \nonumber \\ \frac{1}{2} \int^{\infty}_{0} r'_{ \perp} d r'_{ \perp}[ \ln \frac{r^{'2}_{\perp }   \gamma p_z}{L_z} - \frac{i \pi}{2}] V(r'_{\perp}, 0) \Psi_0(0, r'_{ \perp}),
\nonumber
\end{eqnarray}
and we obtain 
\begin{eqnarray}
|\Psi_0(0,0)|^2=\frac{1}{B+C}, \label{r21} \\
B=[1 -  \frac{1}{2} \int^{\infty}_{0} r'_{ \perp} d r'_{ \perp} \ln \frac{r^{'2}_{\perp }   \gamma p_z}{L_z} V(r'_{\perp}, 0) ]^2, \nonumber \\
C=[  \frac{\pi}{4} \int^{\infty}_{0} r'_{ \perp} d r'_{ \perp}  V(r'_{\perp}, 0) ]^2.
\nonumber 
\end{eqnarray}
For $m > 0$, we obtain the obvious result $\Psi_m(0,0)= 0$. 
Using the Molière potential [14] as an atomic potential, we 
can easily evaluate (for the corresponding average potential [15])  the integrals which appear in Eq. (21) (see [16]).  We then obtain 
\begin{eqnarray}
|\Psi_0(0,0)|^2=\frac{1}{B_1+C_1}, \label{r22a} \\
B_1=[1 -  M \frac{2 Z_1 Z_2 e^2 R^2}{c} \sum_{i=1}^{3} \alpha_i  \beta_i^{-2} \ln \frac{R^{2}   \gamma p_z}{\beta_i^{2} L_z} ]^2, \nonumber \\
C_1=[  \frac{\pi}{2}  M \frac{2 Z_1 Z_2 e^2 R^2}{c} \sum_{i=1}^{3} \alpha_i  \beta_i^{-2}  ]^2,
\nonumber 
\label{r22b}
\end{eqnarray}
where $Z_1$ is the charge of an atom in the target and $Z_2$ is 
the charge of the incident partic1e; $\alpha_i$ and $ \beta_i$ are the parameters of the Molière potential. It follows from Eq. (22) that the density of positively charged particles on the 
axis of an extended chain ($L_z >> p_z R^2$) is always lower than 
unity and the density of negatively charged partic1es can 
exceed unity.

We note that the proposed approximation holds for an  arbitrary strength of the potential $U_0 = 2 Z_1  Z_2 e^2 /c$.  ln particular, for $U_0 \equiv 0$, we obtain $|\Psi_0(0,0)|^2= 1$ which is the 
correct result in the absence of scatterers. The dependence of  $|\Psi_0(0,0)|^2$ on $U_0$ is shown in Fig. 1. The locations and magnitudes of the maxima of the curves in Fig. 1 can be easily determined from Eq. (22) 

\begin{eqnarray}
U_{0 max}= \nonumber \\
\frac{ \sum_{i=1}^{3} \alpha_i  \beta_i^{-2} \ln \frac{R^{2}   \gamma p_z}{\beta_i^{2} L_z}}{ M  R^2([ \sum_{i=1}^{3} \alpha_i  \beta_i^{-2} \ln \frac{R^{2}   \gamma p_z}{\beta_i^{2} L_z}]^2+[ \frac{\pi}{2} \sum_{i=1}^{3} \alpha_i  \beta_i^{-2} ]^2)} , \label{r23a} \\
|\Psi_{0 max}(0,0)|^2=\frac{1}{B_1+C_1}, \nonumber \\
B_1=[1 -  M  R^2 U_{0 max}  \sum_{i=1}^{3} \alpha_i  \beta_i^{-2} \ln \frac{R^{2}   \gamma p_z}{\beta_i^{2} L_z} ]^2, \nonumber \\
C_1=[  \frac{\pi}{2} M  R^2 U_{0 max}  \sum_{i=1}^{3} \alpha_i  \beta_i^{-2}  ]^2.
\nonumber
\end{eqnarray}
For $| \ln(\gamma R^2 p_z/L_z) |>>1$, the density of negatively charged 
particles on the axis of the chain is logarithmically large and independent of the magnitude of the potential 
\begin{eqnarray}
|\Psi_{0 max}(0,0)|^2=  (\frac{\pi}{2} )^2 \frac{ \sum_{i=1}^{3} \alpha_i  \beta_i^{-2} \ln \frac{R^{2}   \gamma p_z}{\beta_i^{2} L_z} }{  \sum_{i=1}^{3} \alpha_i  \beta_i^{-2} }
\label{r24}
\end{eqnarray}
The radial distribution of charged partlc1es across the 
potential of the chain, obtained in this case, is shown in 
Fig. 2. For real crystals, the density on the chain axis 
can increase only for low-energy electrons. For other 
negatively charged partic1es ($\mu^{-}$, $\pi^{-}$, $\bar{p}$, etc.) which have 
large masses and for ultrarelativistic electrons $e^{-}$ [when 
$M$  in Eq. (3) is replaced by $E$], the quantity $U_{0 max}$ becomes negligibly small for $\ln(L_z/\gamma p_z R^2)>>1$ (see Table 1 where the calculations are presented for  $\ln(L_z/\gamma p_z R^2)=10$). 
To observe an appreciable increase in the density 
for $\mu^{-}$, $\pi^{-}$, and $\bar{p}$, it would be necessary to use high-index 
crystallographic directions but the effect of neighboring atomic chains would then be important.

\begin{table}[htbp]
\caption{table title}
\begin{tabular}{|p{1.3cm}|p{1.3cm}|p{1.3cm}|p{1.3cm}|p{1.3cm}|}
\hline
Crystal & $e^{-}$ & $\mu^{-}$ & $\pi^{-}$ & $\bar{p} $ \\
\hline

&&\\[-1ex]
Be
& $2.6$
& $0.013$
& $0.009$
& $0.0014$
\\

&&\\[-1ex]
C
& $3.4$
& $0.016$
& $0.012$
& $0.0019$
\\

&&\\[-1ex]
Si
& $6.0$
& $0.029$
& $0.022$
& $0.0033$
\\

&&\\[-1ex]
Ge
& $10.4$
& $0.05$
& $0.038$
& $0.0057$
\\

\hline

\end{tabular}
\end{table}

We shall now derive expressions for the amplitude 
and total scattering cross section. The wave function 
defined by Eq. (9) is valid only in the region $|{\bf r_z} |\lesssim L_z$ 
(see Sec. 3). It follows that it can be used to calculate 
the amplitude provided Eq. (9) is substituted in the expression 
\begin{eqnarray}
 f=-\frac{1}{4 \pi} \int d {\bf r} e^{-i {\bf p_{f} r}} V({\bf r}) \Psi({\bf r})  
\label{r25}
\end{eqnarray}
which requires the knowledge of the wave function only 
in the region of action of the potential. We stress that 
such situation is analogous to the situation in the eikonal 
approximation. The eikonal wave function is a good approximation only in the region of space $|{\bf r_z}|<< p_z R^2$  (see [17]) and its asymptotic behavior for $|{\bf r_z}| \rightarrow \infty$ cannot 
be used to calculate the scattering amplitude. Substituting Eq. (9) in Eq. (25), we obtain 
\begin{eqnarray}
 f=-\frac{1}{4 \pi} \frac{\sin(N_s(p_{iz}-p_{f z})\frac{c}{2})}{\sin((p_{iz}-p_{f z})\frac{c}{2})} c \times \nonumber \\ \int d {\bf r_{\perp}} e^{-i {\bf p_{\perp f} r_{\perp }}} V_{\perp}({\bf  r_{\perp}}, p_{i z}- p_{f z}) \psi_{\bf{p} i \perp } ({\bf r_{\perp }}) , 
\label{r26}
\end{eqnarray}
where $V_{\perp}({\bf  r_{\perp}}, q)$ is the longitudinal component of the chain 
Fourier potential $V_{\perp}({\bf  r_{\perp}}, q) =V_{at}({\bf  r_{\perp}}, q) /c$. Since the integral in Eq. (26) is a slowly varying function of $ p_{i z}- p_{f z}$ compared with the function outside the integral, we can 
write the scattering amplitude in the form 
\begin{eqnarray}
f=-\frac{1}{2 \pi} \frac{\sin((p_{iz}-p_{f z})\frac{L_z}{2})}{(p_{iz}-p_{f z})}  \times \nonumber \\ \int d {\bf r_{\perp}} e^{-i {\bf p_{\perp f} r_{\perp }}} V({\bf  r_{\perp}},0) \psi_{\bf{p} i \perp } ({\bf r_{\perp }}) . 
\label{r27}
\end{eqnarray}
For small angles of incidence $\theta_0 <<1$ and for small angles 
of partic1es leaving the crystal  $\theta_1 <<1$  measured from the 
direction of the Oz axis, we obtain 
\begin{eqnarray}
p_{i z}- p_{f z}=2p \sin(\frac{\theta_0+ \theta_1}{2}) \sin(\frac{\theta_1- \theta_0}{2}) \simeq \frac{1}{2} p_z(\theta_1^2- \theta_0^2) 
\nonumber
\end{eqnarray}
It follows that the amplitude defined by Eq. (27) and the 
differential scattering cross section $d \sigma /d \Omega = |f|^2$ have 
sharp maxima at  $\theta_0  =\theta_1 $. For angles of incidence  $\theta_0 >> 2/ \sqrt{p_z L_z}$,  the polar width of the maximum for an extended 
potential  $L_z >>   p_z R^2 $  is given by 
\begin{eqnarray}
|\theta_1- \theta_0|_{ef} \sim  \simeq \frac{2}{p_z L_z \theta_0} .
\nonumber
\end{eqnarray}
The spatial distribution of the amplitudes defined by Eq. 
(27) is then either ring- or doughnut-shaped, which is 
observed in experiments on hyperchanneling [18 ,19].  For 
$ \theta_0 \lesssim  1/ \sqrt{p_z L_z}$, the ring is compressed and becomes a 
peak. For $ \theta_0 << 2/ \sqrt{p_z L_z}$, the width of the peak is given 
by 
\begin{eqnarray}
|\theta_1- \theta_0|_{eff}  \sim  \simeq \frac{2}{p_z L_z } 
\nonumber
\end{eqnarray}
Using the optical theorem $ \sigma = (4 \pi / p) Im \;f(0)$, we obtain the following result for the total scattering cross section: 
\begin{eqnarray}
 \sigma=-\frac{ L_z }{p}  Im  \int d {\bf r_{\perp}} e^{-i {\bf p_{\perp f} r_{\perp }}} V({\bf  r_{\perp}},0) \psi_{p i \perp } ({\bf r_{\perp }})  
\label{r28}
\end{eqnarray}
For small angles of incidence $ \theta_0 << 2/ \sqrt{p_z L_z}$, the cross 
section can be approximated by 
\begin{eqnarray}
 \sigma=\frac{ L_z }{8 p} \left[ \left (\frac{1-\frac{1}{2} \int  r_{\perp} d  r_{\perp}  V(  r_{\perp},0)  \ln\frac{  r^2_{\perp} \gamma p_z}{L_z}}{2 \pi \int r_{\perp} d  r_{\perp}  V(  r_{\perp},0) }\right )^2  + \frac{1}{64}  \right ]^{-1}. \;\;
\label{r29}
\end{eqnarray}
With an accuracy up to a coefficient depending on the 
form of the potential, Eq. (29) reduces to the results of 
model calculations [20,21].  For angles of incidence  $ \theta_0 << 2/ \sqrt{p_z L_z}$, the wave function $ \psi_{\bf{p} i \perp } ({\bf r_{\perp }})$ is determined by 
the two-dimensional Lippmann-Schwinger equation and 
the cross section defined by Eq. (28) is proportional to 
the two-dimensional scattering amplitude from a potential $V_{\perp}({\bf  r_{\perp}}, 0)$ which may be resonant for negatively charged particles [21].

\section{3. SCATTERING FROM AN EXTENDED ATOMIC PLANE }
\label{sec:scattering2}

We shall now consider scattering from the potential 
of a rectangular atomic plane containing $N_y$ atomic chains 
separated by a period $b$ and each containing $N_z$ atoms. 
For simplicity, we shall assume that $N_y$ and $N_z$ are odd 
numbers
\begin{eqnarray}
U({\bf r_{\perp}})=\sum_{-(N_z-1/2)}^{N_z-1/2} \sum_{-(N_y-1/2)}^{N_y-1/2}  U_{at} ({\bf r}+{\bf n_z}  c j_z +{\bf n_y}  b j_y ), \nonumber
 \\
U({\bf k})= U_{at} ({\bf k})  \frac{\sin(N_z k_z \frac{c}{2})}{\sin(k_z \frac{c}{2})}  \frac{\sin(N_y k_y \frac{b}{2})}{\sin(k_y \frac{b}{2})}. \;\;
\label{r30}
\end{eqnarray}
As in the case of the chain potential, the eikonal approximation is not valid for an extended plane potential $L_z >> 
p_z R^2$ and $L_y >> p_y R^2$. We shall, therefore, base our discussion on the approach developed in Sec. 1 for scattering from a chain and seek the expansion of Eq. (5) for $L_z = 
N_z c\rightarrow \infty$ and $L_y = N_y b\rightarrow \infty$,  and then perform expansion 
for $p_z  \rightarrow \infty$ and  $p_y  \rightarrow \infty$. The final result which holds subject to the additional condition $p_y/ p_z \rightarrow  0$ has the form 
\begin{eqnarray}
  \Psi^{(1)}({\bf r}) =e^{i {\bf p r}} \int^{\infty}_{-\infty} d R_{x} (-\frac{i}{2})[\frac{e^{i p_x |R_x|}}{p_x}-  i\left(\frac{\pi |R_x|}{2 p_x}\right)^{1/2} \times \; \label{r31} \\ 
   \Phi_{1/2}(\frac{i \pi}{2} -\ln \frac{L_z  p_{x}}{2 p_z |R_{x}|}, p_{x} |R_{x}|)	]  e^{- i p_{x} R_{x}}  V_{\perp}({r_{x} -R_{x} },  0,0). \nonumber
\end{eqnarray}
Here, $ \Phi_{1/2}(\beta, \rho)$ is the incomplete cylindrical function 
of fractional order [13] and $ V_{\perp}(r_{x},  0,0)= V_{at}(r_{x} ,  0,0)/c b$ 
is the average potential of an atomic plane. Since the integral in Eq. (31) is independent of $r_y$ and $r_z$ , we can perform expansions in the multiple integral which determines 
the correction $\Psi^{(n)}$ as in the case of an atomic chain, 
which yields 
\begin{eqnarray}
\Psi^{(n)}({\bf r}) =e^{i {\bf p r}} \int^{\infty}_{-\infty} d R_{1 x} ...\int^{\infty}_{-\infty} d R_{n  x} \tilde{G}_{p_{x}} ( R_{x})  e^{- i p_x R_{1x}}   \nonumber  \\ \times V_{\perp} ( r_{x} -R_{1 x }, 0,0) 
	 ... \tilde{G}_{p_{ x}}( R_{n  x}) e^{- i p_x R_{n x}} \nonumber \\ \times V_{\perp} ( r_{x} - R_{1 x} -... - R_{n x} , 0 ,0),  \;\;  
\label{r32}
\end{eqnarray}

\begin{eqnarray}
\tilde{G}_{p_{x}}( R_{x})= (-\frac{i}{2})[\frac{e^{i p_x |R_x|}}{p_x}-  i\left(\frac{\pi |R_x|}{2 p_x}\right)^{1/2} \times \; \nonumber \\ 
   \Phi_{1/2}(\frac{i \pi}{2} -\ln \frac{L_z  p_{x}}{2 p_z |R_{x}|}, p_{x} |R_{x}|)	] .
\label{r33}
\end{eqnarray}
Summation of the series in Eq. (4) with the terms defined 
by Eq. (32) yields the following expression for the wave 
function in the field of a rectangular atomic plane: 
\begin{eqnarray}
 \Psi({\bf r})=e^{i p_z r_z+i p_y r_y} \phi_{p_x} (r_x), 
\label{r34}
\end{eqnarray}
where the transverse function $ \phi_{p_x} (r_x)$ satisfies an integral equation 
\begin{eqnarray}
  \phi_{p_x} (r_x)=e^{i p_x r_x} +  \int^{\infty}_{-\infty} d r'_{ x} \tilde{G}_{p_{x}}( |r_{x}-r'_x|) \times \nonumber \\
		V_{\perp} ( r'_{x} , 0,0)  \phi_{p_x} (r'_x), 
\label{r35}
\end{eqnarray}
with the kernel $ \tilde{G}_{p_{x}}( R_{x}) $ determined by Eq. (33). 

The distribution of the probability density, the amplitude, and the scattering cross section in the field of an 
atomic plane can be obtained from Eq. (35) by the method described for the atomic chain.

\section{4. DISCUSSION OF RESULTS }
\label{sec:scattering3}

The conditions of validity of the present approximation can be obtained from the requirement that the terms 
which were neglected in the derivation of Eqs. (10) and 
(35) should be small compared with the terms retained, i.e., 
\begin{eqnarray}
|\Psi ^{(0)}|  >> |\Psi_{m} ^{(n)}| , \; \;|\Psi _{1}^{(n)}|  >> |\Psi_{m} ^{(n)}| , \; m \geq 2,
\label{r36}
\end{eqnarray}
where $\Psi_m^{(n)}$ are the terms in the asymptotic expansion 
of the wave function  $\Psi^{(n)}$ of the Born series defined by 
Eq. (4). We can write $\Psi^{(1)}$ with an accuracy up to terms 
$O(1/p^2_z + 1/p_z L_z + 1/ L_z^2)$ for an atomic chain 
\begin{eqnarray}
 \Psi^{(1)} =\Psi_1^{(1)}+\Psi_2^{(1)}+\Psi_3^{(1)},\;\;\;\;  \label{r37}  \\ 
 \Psi_1^{(1)}({\bf r}) =e^{i {\bf p r}} \int d {\bf R_{ \perp}} \left(-\frac{i}{8}\right) [H_0^{(1)}(p_{\perp} R_{ \perp}) + \;\;  \nonumber \\
 H_0( \ln \frac{N_z c p_{\perp}}{2 p_z R_{ \perp}}, p_{\perp} R_{ \perp})	]  e^{- i{\bf p_{\perp} R_{ \perp}}}  V_{\perp}({\bf r_{\perp} -R_{ \perp} },  0), \;\;  \nonumber \\
 \Psi_2^{(1)}({\bf r}) =e^{i {\bf p r}}  \frac{r_z}{2 \pi L_z}  \int d {\bf R_{ \perp}}  e^{ i \frac{p_z}{L_z} ({\bf R_{ \perp}}- {\bf p_{\perp}} \frac{L_z}{2 p_z})^2} \times \nonumber \\  V_{\perp}({\bf r_{\perp} -R_{ \perp} },  0), \;\;  \nonumber \\
 \Psi_3^{(1)}({\bf r}) =e^{i {\bf p r}} \sum_{n=\pm 1,\pm 2,...} \frac{(-1)^{(N_z-1)n} c }{4 \pi p_z n} e^{ i \frac{2 \pi n}{c}r_z} \times \nonumber \\  V_{\perp}({\bf r_{\perp} }, \frac{2 \pi n}{c}) .\;\;  \nonumber 
\end{eqnarray}

(a) We shall now consider small angles of incidence 
 $ \theta_0 << 2/ \sqrt{p_z L_z}$. The integrals in Eq. (37) can be easily 
estimated 
\begin{eqnarray}
|\Psi_1^{(1)} | \simeq  | \bar{V}_{\perp} R^2 \ln \frac{p_z \gamma R^2}{L_z} |, \; |\Psi_2^{(1)} | \simeq  | \bar{V}_{\perp} R^2  \frac{r_z }{2 \pi L_z} |, \; \nonumber \\
|\Psi_3^{(1)} | \simeq  | \bar{V}_{\perp} \frac{c }{4 \pi p_z} |.
\nonumber
\end{eqnarray}
Comparing $|\Psi_1^{(1)} |$ with $|\Psi_2^{(1)} |$, we find that the wave function defined by Eq. (9) is valid in the region 
\begin{eqnarray}
|r_z| <<  |2 \pi L_z \ln \frac{L_z}{p_z \gamma R^2} |.
\label{r38}
\end{eqnarray}
i.e., for $ L_z >> p_z R^2$, the wave function defined by Eq. (9) 
is valid virtually in the whole region of action of the chain 
potential and can be used in the calculation of the scattering amplitude as described above. Comparing $|\Psi_1^{(1)} |$  and   $|\Psi_3^{(1)} |$, we obtain the following important criterion: 
\begin{eqnarray}
|p_z R^2 \ln \frac{L_z}{p_z \gamma R^2} | >> \frac{c}{4 \pi}
\label{r39}
\end{eqnarray}
which indicates that the dynamic length of longitudinal 
coherence $p_z R^2$ should exceed the distance $c$ separating 
the atom under study from the preceding and following 
atoms. We then obtain physical averaging of the potential 
of an atomic chain.

(b) We shall now assume that the angle of incidence 
satisfies  $ \theta_0 \gtrsim 2/ \sqrt{p_z L_z}$. We then obtain 
\begin{eqnarray}
|\Psi_1^{(1)} |  \simeq  | \frac{\bar{V}_{\perp} R^2}{\sqrt{p_{\perp} R}} |, \;\;\; |\Psi_2^{(1)} | \simeq  | \frac{n_z \bar{V}_{\perp} R^2}{2 \pi L_z} |.
\nonumber
\end{eqnarray}
The condition $|\Psi_1^{(1)} | >> |\Psi_2^{(1)} | $ yields the following restriction on the angle of incidence: 
\begin{eqnarray}
\theta_0 <<\frac{ (2 \pi )^2 }{p_z R} .
\label{r40}
\end{eqnarray}

To derive this inequality, we set $r_z  \sim  L_z$ to satisfy the 
requirement that the wave function defined by Eq. (9) 
should be valid in the whole range of action of the chain 
potential. Such restriction can be interpreted as follows. 
It is well known that angles  $ \lesssim 1/p_z R$ are important in the 
scattering from an isolated atom. For coherent scattering from the whole chain, we require that the next atom 
should lie within the diffraction cone of the scattering  from the preceding atom. This condition can be satisfied 
provided the axis of the chain lies within a conical diffraction surface. When it lies outside such a conical surface, 
noncoherent effects become important. Finally, the additional requirement $|\Psi_2^{(1)} |,|\Psi_3^{(1)} |<<1$ which follows from the first condition in Eq. (36) should be satisfied. Combining this requirement with the condition (40), we find that the Lindhard angle $\theta_L \sim \sqrt{U_{\perp} / E} $  lies in the range defined by Eq. (40), i.e., the average potential is applicable for particles above the barrier. We wish to stress that 
the condition (40) does not contradict our initial assumption (b) when the condition $ \theta_0 \gtrsim 2/ \sqrt{p_z L_z}$ is satisfied, since, for an extended potential $L_z >> p_z R^2$, the inequality 
$1/p_z R >> 1/ \sqrt{p_z L_z}$ always holds.

The mathematical framework of the present approach  is similar to the eikonal approximation, but our method 
is valid also in the opposite limiting case. It can be easily seen that the conditions derived above follow from the 
conditions of the eikonal approximation [22 ,23] (where the expansion is with respect to $p_z$ rather than with respect to $p$ ) provided $p_z R^2$ is replaced by $L_z$. The eikonal approximation neglects the diffraction effects [17 ,23] In the opposite limit discussed earlier, the diffraction effects 
become dominant since the diameter of the chain $R$ is 
much smaller than the dimensions of the Fresnel zone $ \sqrt{L_z/p_z}$ and the scattering cross section related to diffraction $\sim  L_z/ p_z$ [see Eq. (29)] exceeds considerably the classical cross section $\sim R^2$. Naturally, such effects are observable only if the length determined by multiple scattering from potential fluctuations is much longer than the length of the chain.


\begin{thebibliography}{10}


\bibitem{1}
J. Lindhard 
{\em K. Dan. Vidensk. Selsk. Mat.-Fys. Medd.}, 34, No.14,  (1965).


\bibitem{2}
N.P. Kalashnikov, Coherent Interactions of Charged Pa rtieles in Single 
Crystals [in Russian], Atomizdat, Moscow (1981). 


\bibitem{3}
V.G. Baryshevskii, Channeling Radiation and Reactions in Crystals at High 
Energies [in Russian], Belorussian State University, Minsk (1982). 

\bibitem{4}
A.I. Akhiezer and N.F. Shul'ga, Usp. Fiz. Nauk 137 , 561 (1982) [Sov. 
Phys. usp. 25, 541 (1982)]. 

\bibitem{5}
V.A. Bazyiev and N.K. Zhevago, Usp. Fiz. Nauk 137 ,605 (1982) [Sov. 
Phys. Usp. 25, 565 (1982)]. 


\bibitem{6}
N.P. Kalashnikov and M. N Strikhanov, Kvantovaya Elektronika (Moscow) 
8, 2293 (1981) [Sov. J. Quantum Electronics, 11, 1405 (1981)]. 


\bibitem{7}
M.A. Kumakhov and G. Shirmer, Atomic Collisions in Crystals [in Russianl 
Atomizdat, Moscow (1980). 


\bibitem{8}
 L.I. Schiff, Phys. Rev. 103,  443 (1956). 

\bibitem{9}
N.P. Kalashnikov and V.D. Mur, Yad. Fiz.,!Q, 1117 (1972) [Sov. J. Nucl. 
Phys. 16, 613 (1973)]. 

\bibitem{10}
M.V. Fedoryuk, Method of Steepest Descent [in Russian], Nauka, Moscow 
(1977). 

\bibitem{11}
J. Winp, Proc. Fifth Conf. on Ordinary and Partial Differential Equations, 
Dundee, Scotland, 1978 (ed. by W. N, Everitt), Springer Verlag, Berlin 
(1978), p. 251 [Lecture Notes in Mathematics, Vol. 827]. 

\bibitem{12}
M.A. Leontovich and V.A. Fok, Zh. Eksp. Teor. Fiz. 16,  557 (1946). 

\bibitem{13}
M.M. Agrest and M. Z. Maksimov, Theory of Incomplete Cylindrical 
Functions and their Applications [in Russian], Atomizdat, Moscow (1965). 

\bibitem{14}
G. Molière, Z. Naturforsch. v.A2, 133 (1947); 3, 78 (1948). 

\bibitem{15}
D.S. Gemmel, Rev. Mod. Phys. v.46,  129 (1974). 

\bibitem{16}
I.S. Gradshteyn and I.M. Ryzhik  (eds.), Table of Integrals, Series, and 
Products, Academic Press, NY(1965). 

\bibitem{17}
L.D. Landau and E.M. Lifshitz, Quantum Mechanics: Non-Relativistic 
Theory, 3rd ed., Pergamon Press, Oxford (1977). 

\bibitem{18}
J.S. Rosner, W.M. Gibson, J. A. Golovchenko, A.N. Goland, and H.E. Wegner, Phys. Rev. B18, 1066 (1978). 

\bibitem{19}
S.K. Andersen, O. Fich, M. Nielsen. H.E. Schiott, E. Uggerhoj, C. 
Vraast Thomsen, G. Charpak, G. Petersen, F. Savli, J. P. Ponpon, and P. Siffert;  Nucl. Phys. B167, 1 (1980). 

\bibitem{20}
N.P. Kalashnikov and  M.N. Strikhanov, Zh. Eksp. Teor. Fiz. 69 ,1253 
(1975) [Sov. Phys. JETP 42, 641 (1975)].

\bibitem{21}
N.P. Kalashnikov and G.V. Kovalev,{\em JETP Lett.},  29, 302 (1979) 
[ Pis'ma Zh. Eksp. Teor. Fiz. 29 , 337 (1979)]. 

\bibitem{22}
A.I. Akhiezer, V.F. Boldyshev, and N.F. Shul'ga;  Fiz. Elem. Chastits At. 
yadra 10, 51 (1979) [Sov. J. Part. Nucl. 10,  19 (1979)]. 

\bibitem{23}
Yu. A. Kravtsov and Yu. 1. Orlov, Usp. Fiz . Nauk 132 , 475 (1980) [Sov. 
Phys. Usp. 23 , 750 (1980)].







\end{thebibliography}
\end{document}